# Photonic circuits for laser stabilization with ultra-low-loss and nonlinear resonators

Kaikai Liu[1], John H. Dallyn[2], Grant M. Brodnik[1], Andrei Isichenko[1], Mark W. Harrington[1], Nitesh Chauhan[1], Debapam Bose[1], Paul A. Morton[3], Scott B. Papp[4,5], Ryan O. Behunin[2,6], and Daniel J. Blumenthal[1,*]

[1] Department of Electrical and Computer Engineering, University of California Santa Barbara, Santa Barbara, CA, USA
[2] Department of Applied Physics and Materials Science, Northern Arizona University, Flagstaff, Arizona, USA
[3] Morton Photonics, West Friendship, MD, USA
[4] Department of Physics, University of Colorado Boulder, Boulder, CO, USA
[5] Time and Frequency Division 688, National Institute of Standards and Technology, Boulder, CO, USA
[6] Center for Materials Interfaces in Research and Applications (¡MIRA!), Northern Arizona University, Flagstaff, AZ, USA
*danb@ucsb.edu

# ABSTRACT

Laser-frequency stabilization with on-chip photonic integrated circuits will provide compact, low cost solutions to realize spectrally pure laser sources. Developing high-performance and scalable lasers is critical for applications including quantum photonics, precision navigation and timing, spectroscopy, and high-capacity fiber communications. We demonstrate a significant advance in compact, stabilized lasers to achieve a record low integral emission linewidth and precision carrier stabilization by combining integrated waveguide nonlinear Brillouin and ultra-low loss waveguide reference resonators. Using a pair of 56.4 Million quality factor (Q) $Si_3N_4$ waveguide ring-resonators, we reduce the free running Brillouin laser linewidth by over an order of magnitude to 330 Hz integral linewidth and stabilize the carrier to $6.5\times10^{-13}$ fractional frequency at 8 ms, reaching the cavity-intrinsic thermorefractive noise limit for frequencies down to 80 Hz. This work demonstrates the lowest linewidth and highest carrier stability achieved to date using planar, CMOS compatible photonic integrated resonators, to the best of our knowledge. These results pave the way to transfer stabilized laser technology from the tabletop to the chip-scale. This advance makes possible scaling the number of stabilized lasers and complexity of atomic and molecular experiments as well as reduced sensitivity to environmental disturbances and portable precision atomic, molecular and optical (AMO) solutions.

# INTRODUCTION

Spectrally pure sources are critical for applications that demand low phase noise and high carrier stability, including atomic clocks[1], microwave photonics[2,3], quantum systems[4,5], atomic, molecular and optical (AMO) applications[6–8], and energy efficient coherent fiber communications applications[9]. Table-scale ultra-stable optical cavities are capable of yielding linewidths as low as 10 mHz with a fractional frequency stability of $10^{-15}$ over ~1 second[10–13]. This level of stability requires cavities with a large optical mode volume, athermalized design, environmental isolation, and mitigation of cavity-intrinsic thermal fluctuations[14,15]. In order to make this performance more widely accessible and translate this technology to a wider range of applications, it is critical to implement stable lasers and their reference cavities in photonic integrated waveguide circuits. Such integration also needs to be compatible with other photonic components and with wafer-scale processing in order to bring higher level functionality, lower cost, size and weight, and reduced sensitivity to environmental disturbances[16–18].

There has been progress in bulk-optic miniaturization of optical frequency reference cavities and stabilized lasers. Compact, centimeter-scale whispering gallery mode resonators can achieve impressive laser stabilization with a thermorefractive-noise limited stability of $6\times10^{-14}$ at 100 ms in a thermally and acoustically isolated vacuum enclosure[19] and in an ambient environment reaching $3\times10^{-13}$ at 100 ms[20]. A dual bulk-optic, tapered fiber coupled resonator solution employed a nonlinear micro-disk Brillouin laser and a reference microrod cavity to achieve an 87 Hz integral linewidth[21]. A centimeter-scale fused silica cavity housed in a completely passive vibration-isolating enclosure achieved a remarkable 25 Hz integral linewidth with $7\times10^{-13}$ fractional frequency stability at 300 ms without any vacuum or sophisticated temperature control or isolation[22]. These cavities reduce the integral linewidth by employing a large resonator mode volume to mitigate thermo-optically induced frequency noise, a primary source of phase instability[19,23]. All-waveguide solutions have also made extraordinary advances. For example, an integrated silica spiral waveguide was used to construct a 140 Million quality factor (Q) resonator to stabilize a fibre laser to $4\times10^{-13}$ at 0.4 ms with a 100 Hz effective linewidth[24]. However, such designs utilize deep etched, air clad waveguides that are sensitive to the environment and require hermetic sealing and modifications to CMOS wafer-scale integration processes. Heterogeneous solutions have shown thermorefractive-noise limited performance down to 10 kHz, yet these solutions suffer from frequency instability that broadens the linewidth to several kHz[25].

In this paper, we report a significant advance in photonic integrated laser stabilization, achieved by locking an ultra-high Q nonlinear 1550 nm $Si_3N_4$ waveguide Brillouin laser resonator to an ultra-low loss $Si_3N_4$ waveguide resonator. Using this configuration, we demonstrate an integral 330 Hz linewidth and Allan deviation of $6.5\times10^{-13}$ at 8 ms, the lowest linewidth and highest stability reported for an all-waveguide design to the best of our knowledge. Both integrated resonators are a dilute optical mode $Si_3N_4$ design with large mode-volume and high Q[26]. The intrinsic 56.4 Million Q corresponds to 0.47 dB m$^{-1}$ propagation loss, with a loaded Q of 28.2

Million, a 0.05 dB m$^{-1}$ absorption-limited loss, and a ~2×10$^6$ μm$^3$ mode volume (~27 μm$^2$ optical mode area and roundtrip length of 7.43 cm). The stabilized laser frequency noise is reduced in two spectral bands. High frequency offset from carrier noise (>100 kHz) is reduced in the nonlinear cavity through fundamental linewidth narrowing of stimulated Brillouin scattering (SBS) lasing with a short phonon lifetime, long photon lifetime, high optical Q and large mode-volume[27,28]. The lower frequency offset from carrier noise (<10 kHz) is reduced with the ultra-low loss reference cavity through a combination of narrow linewidth cavity frequency discrimination, and low thermorefractive noise from the large mode volume. We compare the measured stabilized laser frequency noise to the cavity-intrinsic thermorefractive noise limit including the Pound-Drever-Hall (PDH)[29] lock loop noise and simulated photothermal and thermorefractive noise. The stabilized laser emission reaches the thermorefractive noise limit down to 80 Hz frequency offset-from-carrier; this level of performance is achieved with the resonators located in an ambient environment with simple passive vibration damping isolation and without vacuum or thermal isolation. These results demonstrate the potential to bring the performance of bulk optic ultra-stable, high Q, miniaturized resonators to planar all-waveguide, wafer-scale compatible solutions, paving the way towards integrated all-waveguide frequency stabilized lasers for AMO, atomic clocks, microwave photonic, precision spectroscopy, and energy-efficient coherent fiber communications. Potential applications are illustrated in Fig. 1, including an energy efficient, frequency stabilized high capacity coherent wavelength division multiplexed communications transceiver [9,30], neutral atom cooling and probing for optical clocks[1,31,32], and ultra-low phase noise microwave signal generation[2,3].

# RESULTS

### Stabilization with nonlinear and ultra-low loss photonic integrated resonators

The process of laser stabilization using nonlinear and ultra-low loss photonic integrated resonators is illustrated in Fig. 2. The output from a pump laser such as a semiconductor laser, represented as a blue arrow in the upper part of Fig. 2a, has a typical Gaussian spectral density $S_E(\delta\nu)$ shown in the blue curve in the lower part of Fig. 2a. The pump laser is frequency locked to the nonlinear waveguide resonator where the SBS lasing process[26]. While the fundamental linewidth of S1 is drastically compressed relative to the pump[27,28], thermally driven processes lead to slow drifts in the carrier frequency of the SBS emission. This nonlinear SBS process reduces the spectral energy in the Lorentzian wings of the S1 lineshape as illustrated in the $S_E(\delta\nu)$ red curve. A tunable single sideband (SSB), with identical lineshape to S1, is generated using an acousto-optic modulator (AOM) and is frequency locked to the ultra-low loss linear resonator to reduce the close-to-carrier noise energy (Fig. 2b). The spectral energy of the low frequency noise in the SSB is frequency discriminated by the ultra-low loss resonator to produce a narrowed Gaussian lineshape, that is close to Lorentzian, at frequencies near the center carrier frequency (green curve in Fig. 2b). At the mid-frequencies, the lineshape takes on a Gaussian quality as it transitions from the wings towards the center. The resulting stabilized laser lineshape (green curve) can be further understood from the frequency noise spectral density (as illustrated in Fig. 2c offset from carrier). The unstabilized pump ($S_f(\nu)$ blue) has a high frequency white noise floor that is directly related to its fundamental linewidth and more specifically the energy in its lineshape wings. The Brillouin nonlinear resonator reduces this high frequency white noise (i) as illustrated in its frequency noise spectrum (red), ), resulting in a reduction in the wing spectral energy in S1 that can be several orders of magnitude (the curves in Fig. 2c are schematically represented on a logarithmic plot). At mid- to low-frequencies, the emission (S1) is dominated by frequency noise from thermorefractive[33] and photothermal[19] effects in the waveguide resonator that broadens the lineshape, forming the Gaussian-like shape at mid-frequencies, as well as technical noise sources at low frequencies (e.g., vibration and environmental effects) determine the degree of near-carrier linewidth reduction (red). The thermorefractive noise (TRN) contribution is minimized by maximizing the optical mode volume in the ultra-low loss resonator, and stabilization can be improved by increasing the resonator quality factor (Q). Locking the Brillouin output S1 to the ultra-low loss resonator using a high gain PDH loop reduces (stabilizes) the lower frequency noise components (ii) and results in the noise spectrum ($S_f(\nu)$ green). The stabilized laser output will take on the low frequency noise qualities of the ultra-low loss resonator, which are determined predominately by thermorefractive noise and technical noise sources. The resulting stabilized lineshape is characteristic of the green curve in Fig. 2b, with two white noise-like flat frequency regions. By combining nonlinear and ultra-low loss integrated resonators, laser stabilization and linewidth reduction can be achieved.

In this work, the nonlinear Brillouin laser resonator and the ultra-low loss reference cavity are of the same design: a high-aspect ratio silicon nitride ($Si_3N_4$) waveguide core, 7 μm wide by 40 nm thick, designed to maximize mode volume and mitigate losses from sidewall scattering[34,35]. The 11.83 mm resonator radius is selected to satisfy the Brillouin lasing frequency matching condition[26]. Further details of the resonator design are given in Supplementary Section I. The Q factor, mode volume, waveguide design and materials of both resonators are important factors in terms of the Brillouin laser's threshold power and high frequency offset from carrier noise (modified ST linewidth) and the integral linewidth and stability achieved with the ultra-low loss reference cavity. Both the nonlinear Brillouin resonator and ultra-low loss reference resonators have an intrinsic Q of 56.4 Million (0.47 dB m$^{-1}$ propagation loss) and loaded Q of 28.2 Million that are measured using an electro-optic modulation (EOM) sideband technique (see Supplementary Section I). Absorption of optical power within the resonator leads to photothermal heating that can produce thermo-optic shifts in the laser frequency. This source of noise plays a critical role at low frequencies in the reference cavity and hence the stabilized integral linewidth[19]. We measure a 0.05 dB m$^{-1}$ resonator absorption limited loss that accounts for 10.6% of the total loss, using the photothermal absorption technique (see Supplementary Section IV)[36,37]. The nonlinear Brillouin laser resonator is operated with S1 only emission, just below the second order Stokes (S2) threshold, with an on-chip pump power of ~42 mW chosen to minimize the S1 fundamental linewidth[26,28]. This high on-chip pump power in the nonlinear Brillouin cavity combined with the pump laser's relative intensity noise (RIN) induces temperature fluctuations through the absorption heating effect, which results in photothermal noise that broadens the free-running Brillouin laser integral linewidth. As a low optical input power in the ultra-low loss reference cavity for laser stabilization results in the reduced photothermal noise, the PDH lock narrows the Brillouin emission integral linewidth and improves carrier stability, reaching the cavity-intrinsic thermorefractive noise limit.

**Brillouin laser stabilization**

The stabilized laser experimental setup is shown in Fig. 3a. The nonlinear Brillouin and ultra-low loss resonators are mounted on active temperature-controlled stages inside passive enclosures to minimize environmental fluctuations (see Supplementary Section I for the stage setup details). The free-running Brillouin S1 emission is modulated by an AOM driven by a voltage-controlled oscillator (VCO) whose output is PDH locked to the reference cavity with an on-chip power of ~0.1 mW and a lock loop bandwidth of ~20 kHz. An unbalanced fiber Mach-Zehnder interferometer (MZI) with a 1.026 MHz free spectral range is used as an optical frequency discriminator (OFD) to measure the frequency noise above ~1 kHz offset from carrier[26,38]. For frequency noise below ~1 kHz frequency offset from carrier, we employ a Rock™ single frequency fiber laser that is PDH locked to a Stable Laser Systems™ ultra-low-expansion (ULE) high-finesse cavity in order to produce a Hz-level linewidth output at 1550 nm wavelength, with a frequency drift of ~0.1 Hz s$^{-1}$ and a frequency stability of ~$10^{-15}$ at 1 s averaging time. We refer to this laser as the stable reference laser (SRL) and the frequency noise measurement using this laser as the SRL frequency noise measurement. The ultra-low loss reference cavity stabilized Brillouin emission is photomixed with the SRL output in a high-speed photodetector to produce a heterodyne signal that carries the Brillouin laser's frequency noise. The resulting heterodyne signal is a ~100 MHz heterodyne beatnote that is measured using a Keysight 53230A frequency counter with a frequency noise floor characterized to be below $10^{-3}$ Hz$^2$ Hz$^{-1}$. With the SRL's ultra-high frequency stability below ~1 kHz offset and below the $10^{-3}$ Hz$^2$ Hz$^{-1}$ frequency noise floor of the frequency counter, we are able to accurately measure both the free-running and stabilized Brillouin laser emission at frequencies below ~1 kHz. The OFD and SRL frequency noise measurements and their limitations are discussed in detail in the Supplementary Section II. The stabilized laser frequency noise measurements are shown in Fig. 3a yielding the free-running and stabilized Brillouin laser fundamental linewidths of 0.72 Hz and 1.57 Hz, respectively.

The increase in the fundamental linewidth in the AOM modulated S1 emission is due to the added AOM frequency noise at high offset frequencies (i.e., the laser fundamental linewidth, high frequency offset from carrier, is better than that of the AOM and its voltage-controller oscillator). The integral linewidths are calculated from the frequency noise spectrum for the free-running and stabilized Brillouin laser, yielding 3.24 kHz and 292 Hz, respectively, a factor of 10 decrease (see Supplementary Section II for the integral linewidth calculation). Taking the Fourier transform of the SRL and the Brillouin laser heterodyne beatnote yields linewidths of 2.93 kHz for the free-running and 330 Hz for the stabilized Brillouin laser, as shown in Fig. 4b. These values are in good agreement with our calculated integral linewidths from the frequency noise spectrum[13]. The measured stabilized S1 frequency noise is close to the resonator-intrinsic thermorefractive noise limit and the corresponding Allan deviation limit for frequencies from 80 Hz to 10 kHz, as shown in Fig. 3b and Fig. 4a. The overlapping Allan deviation shown in Fig. 4a is calculated from the time trace of heterodyne frequency recorded by the frequency counter which for the stabilized Brillouin laser reaches a minimum of $6.5\times10^{-13}$ at 8 ms averaging time. The heterodyne beatnote frequency recorded by the frequency counter shown in Fig. 4c with a sampling rate of 1 kHz demonstrates the reduction in the laser frequency fluctuation from the reference cavity stabilization. The thermorefractive noise sets a lower limit for the Allan deviation at short averaging times (below 10 ms) and the low-frequency random walk frequency noise (with an estimated drifting speed of ~10 kHz s$^{-1}$) increases the Allan deviation at longer averaging times (above 10 ms), represented by the purple-dashed curve in Fig. 4a. This drift results from the long-term environmental temperature drift.

## Frequency noise modeling

We incorporate our frequency noise measurements into modeling to determine the role that the nonlinear Brillouin and ultra-low loss reference cavities play in laser stabilization and what is the primary frequency noise contribution. The noise sources include shot, photodetector, photothermal and thermorefractive noise (see Fig. 3b). Shot and detector noise are below the thermorefractive noise limit calculated using Ref. [33] (Fig. 3b). The thermorefractive noise is the same for both the nonlinear and reference resonators, as both resonators have the same physical parameters, while the photothermal noise is closely dependent on the on-chip optical power and optical power fluctuations. In order to model the photothermal noise, we generalize the analysis of Ref. [33] to include spatial changes in the temperature field produced by fluctuations in absorbed power (Supplementary Section IV). The inputs to our frequency noise model include the pump laser and Brillouin laser RIN (shown in Supplementary Section IV), the on-chip free running Brillouin laser input pump power, the reference cavity on-chip input power, the resonator optical absorption loss as measured by the photothermal technique (see Supplementary Section IV), the cavity build-up factor and optical mode profiles. Our modelling shows that the free-running Brillouin laser frequency noise at low frequency offset (below 10 kHz) is dominated by photothermal noise (orange-dashed curve in Fig. 3b) and that the ultra-low loss reference cavity's frequency noise is limited by the thermorefractive noise (green-dashed curve in Fig. 3b) at frequency offset above 10 Hz, so that the thermorefractive noise sets the fundamental limit for the stabilized laser at frequency offset below 10 kHz.

The dominant contributions to thermally driven frequency instability consist of two primary components; thermorefractive noise, that arises from fundamental thermodynamic fluctuations within both the nonlinear Brillouin and ultra-low loss reference resonators, and photothermal noise, produced by thermal fluctuations driven by power absorbed from a fluctuating optical field (Fig. 3b)[33]. When the on-chip optical powers within the nonlinear Brillouin laser and ultra-low loss reference resonators are different, the photothermal noise induced by the optical power fluctuation and the high on-chip pump power in the nonlinear resonator can rise above the thermorefractive noise and the photothermal noise in the reference resonator falls below the thermorefractive noise.

Measurements of the photothermal noise are shown in Fig. 3c, where the on-chip input power in the ultra-low loss reference cavity is increased from 0, to 7 and 10 dBm, and the frequency noise below 10 kHz of the stabilized Brillouin laser measured by the heterodyne beatnote with the SRL increases accordingly. In Fig. 3c, the dashed curves are the modelled photothermal noise with an estimated 5, 8 and 12 dBm on-chip input power. The difference in the experimentally recorded on-chip power and the on-chip power used in the photothermal noise estimation could result from calibration accuracy of the on-chip power. Since the photothermal noise is proportional to the square of the on-chip input power, this observation confirms that the photothermal noise in the ultra-low loss reference cavity stabilization with ~0.1 mW input power falls below the thermorefractive noise limit, which is consistent with the conclusion draw from our photothermal noise modeling.

## DISCUSSION

We report a photonic integrated stabilized laser with a record low 330 Hz integral linewidth and frequency stability of $6.5 \times 10^{-13}$ at 8 ms, the lowest linewidth and highest laser stability demonstrated for all-waveguide devices, to the best of our knowledge. The nonlinear and reference cavity configuration employs identical 56.4 Million intrinsic Q photonic integrated bus-coupled ring resonators. Measurements and simulations of the cavity frequency noise dynamics, including the cavity-intrinsic thermorefractive noise, the on-chip power fluctuation induced photothermal noise, as well as the technical noise of the PDH locking loop, show that laser emission stabilization is limited by the cavity intrinsic thermorefractive noise for frequencies as low as 80 Hz, while the ambient environmental noise such as the thermal drift becomes dominant at frequencies below 80 Hz. While the Brillouin laser nonlinear cavity is driven by photothermal noise, the identical high Q reference is able to stabilize the free-running Brillouin laser down to its thermorefractive noise limit, and the high frequency offset from carrier noise is determined by the Brillouin laser. This approach shows that frequency noise bandwidth engineering can be achieved by combining the underlying limits of the nonlinear Brillouin and ultra-low loss reference cavities in different frequency ranges as well as reducing noise.

The photothermal and technical noise in the lock loop and on-chip input power to the reference cavity are critical factors to performance, and the Q factor is critical for both the Brillouin lasing emission and laser stabilization. Since the Brillouin lasing threshold is inversely proportional to $Q^2$, a higher Q lowers the Brillouin laser threshold and intracavity power fluctuations, leading to the reduced photothermal noise. Under cavity locking, the Q determines the signal-to-noise ratio in the frequency discrimination of the laser frequency against the cavity resonance, which is proportional to Q and the optical input power. A high Q factor is required to tightly lock the nonlinear cavity output to the reference cavity with a small optical input power, such that the PDH lock is not limited by technical noise or the photothermal noise induced by the optical input power fluctuation. Under these conditions, laser stabilization performance is only limited by the reference cavity itself and the environmental noise. While identical in geometry and optical properties, there are differences in the thermally driven noise in the nonlinear and reference cavities due to intra-cavity power fluctuations and photothermal noise levels. This frequency noise difference in the nonlinear and reference cavities can be decreased in the future, as the Brillouin laser threshold and the pump power is reduced with increased cavity Qs, so that the low pump power leads to low intra-cavity power fluctuations and reduced photothermal noise. Future improvements include increasing the mode volume for a lower thermorefractive noise floor by using a coil resonator with a long roundtrip close-loop coil as the cavity and a low propagation loss for high Qs to further suppress the laser noise at frequencies from ~100 Hz to ~100 kHz. These results show promise to bring the performance of ultra-high Q resonators to CMOS compatible stabilized laser technology, and pave the way to scale the number of stabilized lasers and complexity for atomic and molecular scientific experiments as well as to reduce sensitivity to environmental disturbances and enable portable solutions to precision applications.

# DATA AVAILABILITY

The data supporting the findings of this study are available within the article and its Supplementary Information. Extra data are available from the corresponding authors on reasonable request

# ACKNOWLEDGEMENTS

This material is supported under ***OPEN 2018*** Advanced Research Projects Agency-Energy (ARPA-E), U.S. Department of Energy, under Award Number DE-AR0001042. The views and conclusions contained in this document are those of the authors and should not be interpreted as representing official policies of ARPA-E or the U.S. Government or any agency thereof. Andrei Isichenko acknowledges the support from the National Defense Science and Engineering Graduate (NDSEG) Fellowship Program.

# CONTRIBUTIONS

K. L., and D. J. B. prepared the manuscript. D. J. B. conceived the dual-resonator Brillouin laser stabilization approach. G. M. B., and K. L. implemented the Brillouin laser stabilization experiment. J. D., and K. L. studied and modelled the photothermal noise and thermorefractive noise. P. A. M. contributed the narrow linewidth integrated optical pump sources. A. M. I., and M. W. H. helped with the Stable Laser System. D. B. fabricated the $Si_3N_4$ integrated high-Q resonator devices. N. C. helped with the resonator Q and linewidth testing. All authors contributed to analyzing the simulated and experimental results. S. B. P., R. O. B, and D. J. B. supervised and led the scientific collaboration.

# COMPETING INTERESTS

The authors declare no competing interests.

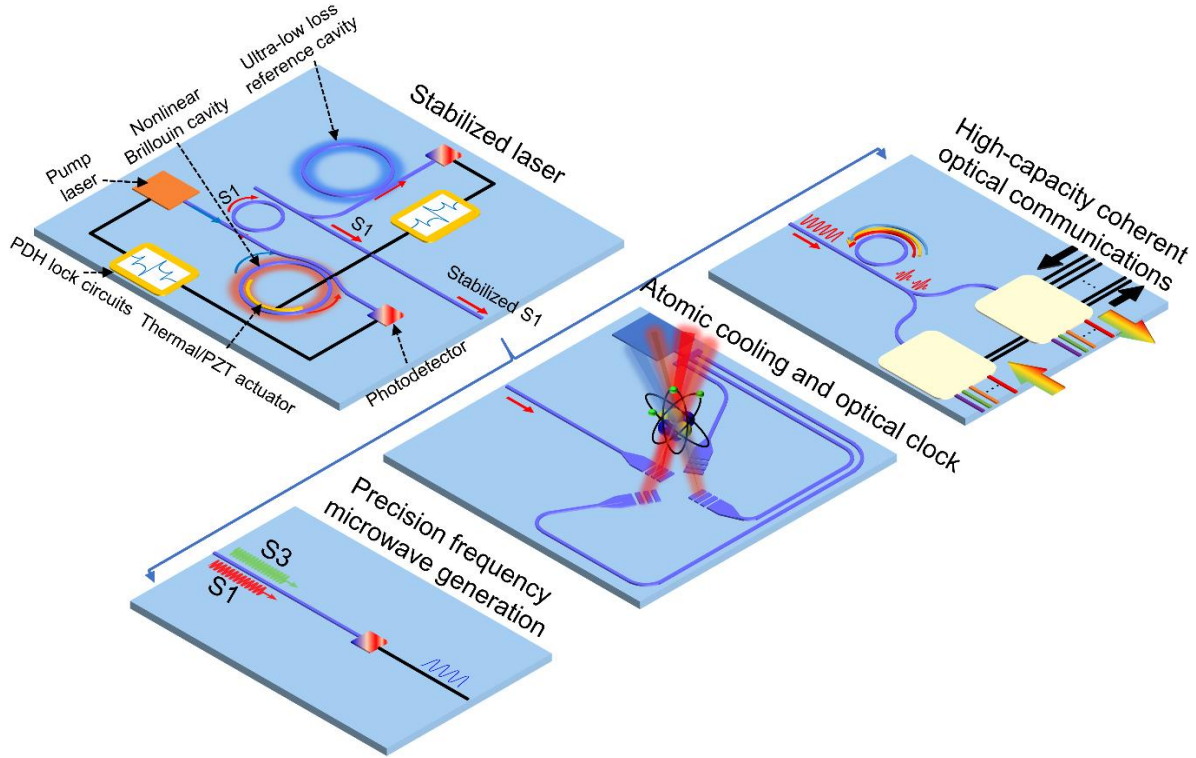

**Fig. 1 | Stabilized laser and its applications.** Stabilized laser generates a spectrally pure emission using the stimulated Brillouin scattering lasing in a nonlinear Brillouin cavity to achieve a low white-frequency-noise floor at high frequency offset from carrier (corresponding to a narrow modified Schawlow-Townes linewidth) and using laser stabilization in a ultra-low loss reference cavity to achieve low frequency noise at low frequency (corresponding to a narrow integral linewidth). Such a stabilized laser chip is advantageous for a wide range of precision applications including high-capacity coherent optical communications, atomic optical clock, and precision frequency microwave generation. PDH, Pound-Drever-Hall. S1, first order Stokes. S3, third order Stokes. PZT, piezoelectric lead zirconate titanate.

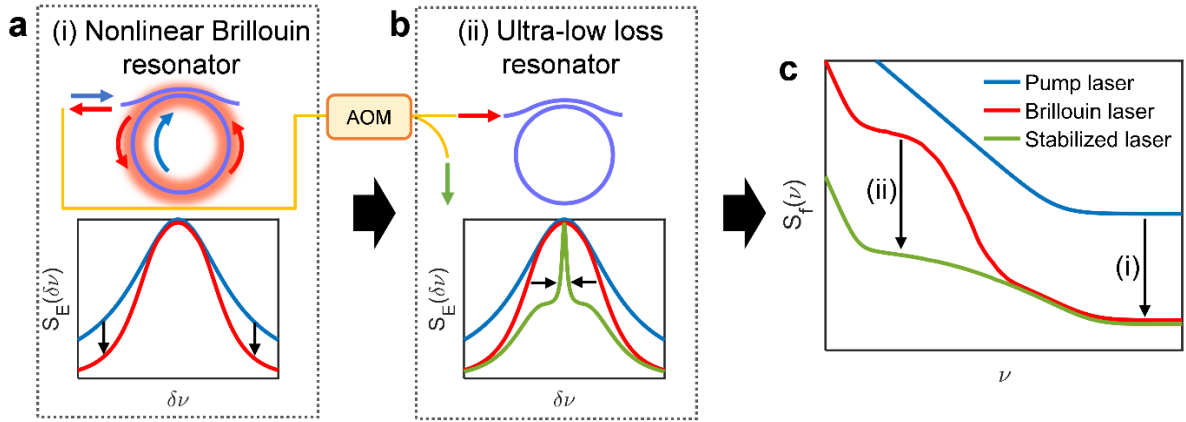

**Fig. 2 | Laser stabilization using nonlinear and ultra-low loss resonators. a**, A pump laser (blue arrow) with Gaussian spectral density $S_E(\delta\nu)$ (blue curve) is frequency locked to a nonlinear resonator that performs Brillouin frequency conversion. The nonlinear Brillouin process reduces the fundamental linewidth of the input pump laser, producing an output first order Stokes (S1) emission (red arrow) that has reduced spectral energy in the Lorentzian wings of the lineshape $S_E(\delta\nu)$ (red curve). **b**, An acousto-optic modulator (AOM) modulates a tunable single sideband onto S1 that is then frequency locked to the ultra-low loss resonator. The low frequency noise spectral energy is filtered out to produce a narrowed lineshape that is close to Lorentzian (green curve) at the center. Gaussian frequency noise energy shapes the lineshape with transition from the low to high frequency noise. **c**, The frequency noise spectral energy (offset from carrier) of an unstabilized pump ($S_f(\nu)$ blue), reduction of the pump high frequency white noise (i) by the nonlinear resonator (red) and reduction of the lower frequency noise components (ii) from the nonlinear resonator output by locking the AOM output to the ultra-low loss cavity yielding the noise spectrum ($S_f(\nu)$ green). The low frequency technical noise (increased noise towards $\nu = 0$) for the green and red curves, will blur out the Lorentzian feature in the lineshape in **b** given a more Gaussian like top as shown.

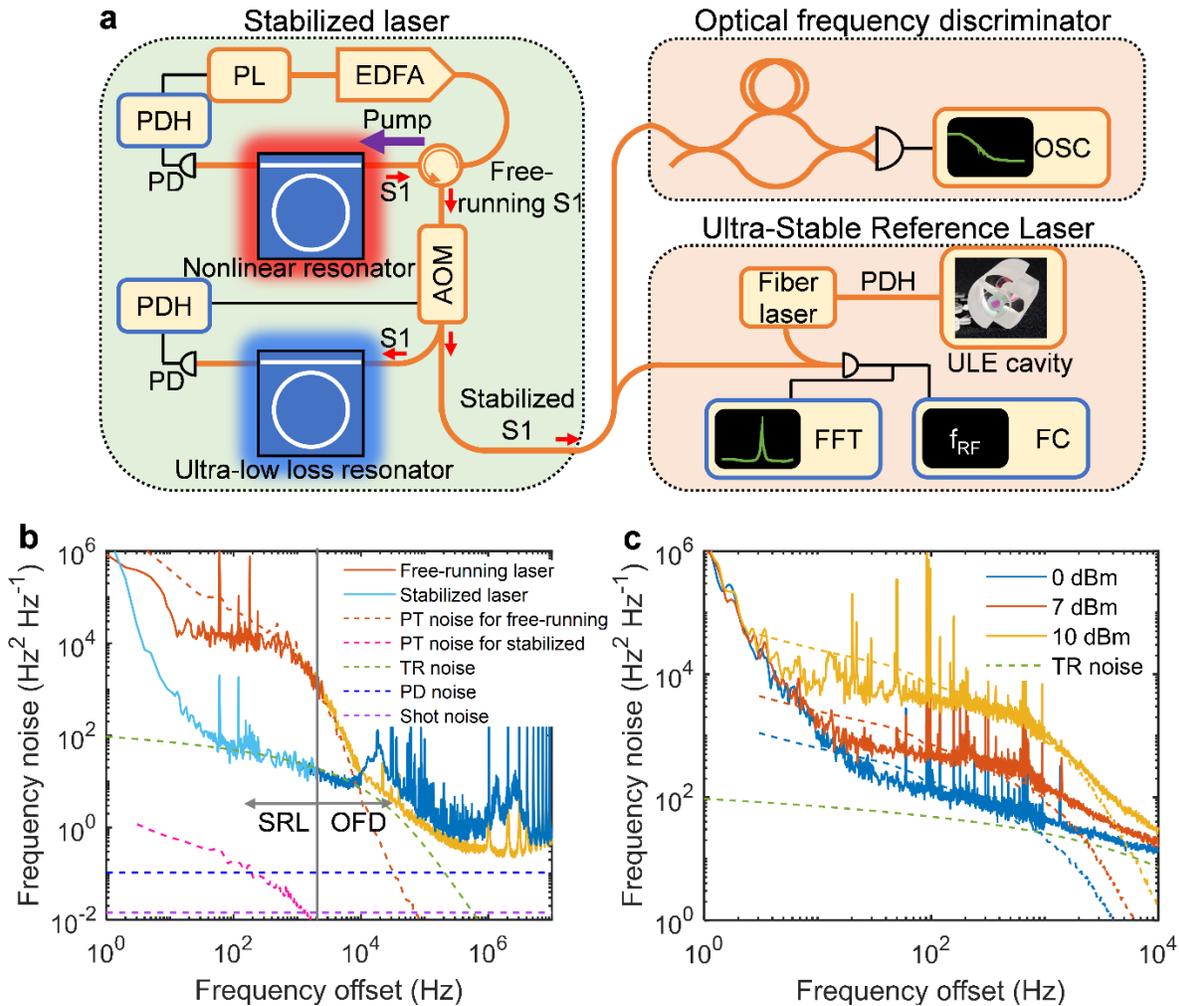

**Fig. 3 | Brillouin laser stabilization and frequency noise measurements and modeling. a**, A pump laser (PL) is Pound-Drever-Hall (PDH) locked to the nonlinear Brillouin cavity to generate Brillouin first order Stokes (S1) emission. The S1 emission out of the reflection port is PDH locked to the ultra-los loss reference cavity using an acousto-optic modulator (AOM). An optical frequency discriminator (OFD) is used to measure the laser's frequency noise above 1kHz frequency offset from carrier, and a heterodyne beatnote from photomixing the Brillouin laser and a stable reference laser (SRL) that is PDH locked to ultra-stable ultra-low-expansion high-finesse cavity is measured at the frequency counter (FC) for frequencies below 1kHz frequency offset from carrier. SRL, stable reference laser; EDFA, Erbium-doped fiber amplifier; ULE, ultra-low expansion; OSC, oscilloscope; PD, photodetector. **b**, OFD and SRL frequency noise measurements for the free-running and stabilized Brillouin laser with the modelled photothermal (PT) noise and thermorefractive (TR) noise in the nonlinear and reference cavities and photodetector (PD) and shot noise in the PDH lock loop. **c**, Frequency noise below 10 kHz offset for the stabilized laser measured by the SRL method increases with the on-chip input power in the reference resonator, as more on-chip input power heats up the resonator and increases the photothermal noise. The on-chip power for the blue, orange and yellow curves are experimentally calibrated to be 0, 7 and 10 dBm. The dashed curves are the corresponding modelled photothermal noise.

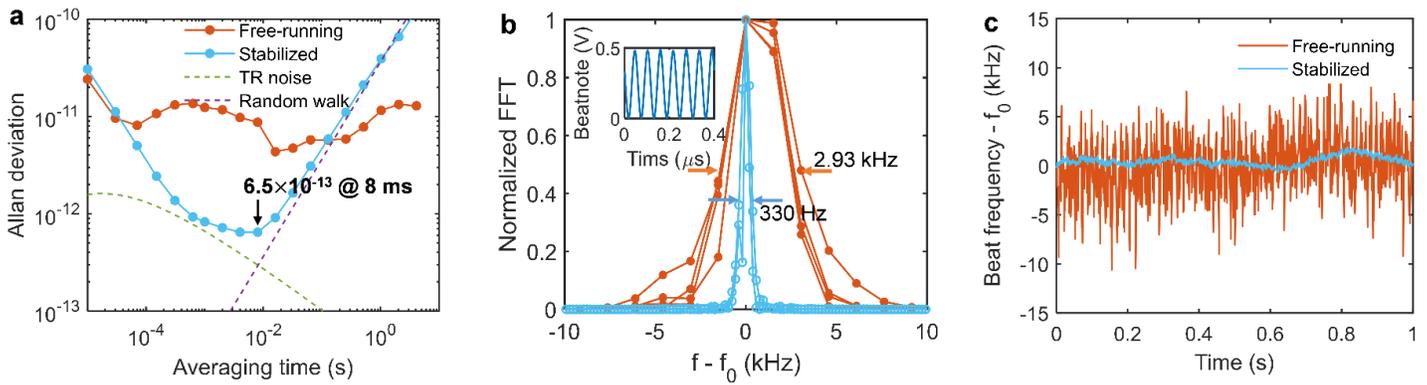

**Fig. 4 | Stabilized laser Allan deviation and linewidth. a**, The overlapping Allan deviation is calculated from the heterodyne beatnote frequency from photomixing the Brillouin laser and the stable reference laser (SLR) and recorded by the frequency counter. **b**, The fast Fourier transform (FFT) of the heterodyne signal reveals the linewidths, with the resolution bandwidth (RBW) of 1 kHz for the orange curve and 100 Hz for the blue curve. **c**, A time trace of the heterodyne beatnote frequency shows the fluctuation of the laser frequency over time, before and after the reference cavity stabilization, where the heterodyne frequency is recorded by the frequency counter with a sampling rate of 1 kHz and the center frequencies ($f_0$) for the orange and blue curves are ~162 MHz and ~156 MHz.

# Photonic circuits for laser stabilization with ultra-low-loss and nonlinear resonators: Supplementary Information

Kaikai Liu[1], John H. Dallyn[2], Grant M. Brodnik[1], Andrei Isichenko[1], Mark W. Harrington[1], Nitesh Chauhan[1], Debapam Bose[1], Paul A. Morton[3], Scott B. Papp[4,5], Ryan O. Behunin[2,6], and Daniel J. Blumenthal[1,*]

[1] Department of Electrical and Computer Engineering, University of California Santa Barbara, Santa Barbara, CA, USA
[2] Department of Applied Physics and Materials Science, Northern Arizona University, Flagstaff, Arizona, USA
[3] Morton Photonics, West Friendship, MD, USA
[4] Department of Physics, University of Colorado Boulder, Boulder, CO, USA
[5] Time and Frequency Division 688, National Institute of Standards and Technology, Boulder, CO, USA
[6] Center for Materials Interfaces in Research and Applications (¡MIRA!), Northern Arizona University, Flagstaff, AZ, USA
*danb@ucsb.edu

## SUPPLEMENTARY INFORMATION

**I. Integrated resonator design and enclosure setup.** The resonator's ring radius is 11.83 mm as shown in Supplementary Fig. 1a, chosen for the frequency matching condition in Brillouin lasing, where 4 times the resonator's free spectral range (FSR) matches the 11.93 GHz Brillouin frequency shift. The bus-to-ring coupling is designed to be critical coupling by choosing a gap of 6 μm. The resonator fabrication processes involve deposition of a 40 nm thick $Si_3N_4$ thin film using low-pressure chemical vapour deposition (LPCVD) onto a 15 μm thermal oxide lower cladding on a silicon base wafer. The waveguides are etched, followed by a 6 μm upper oxide cladding deposition using tetraethoxysilane pre-cursor plasma-enhanced chemical vapour deposition (TEOS-PECVD). A final 11 hours annealing at ~1100 C is performed[1].

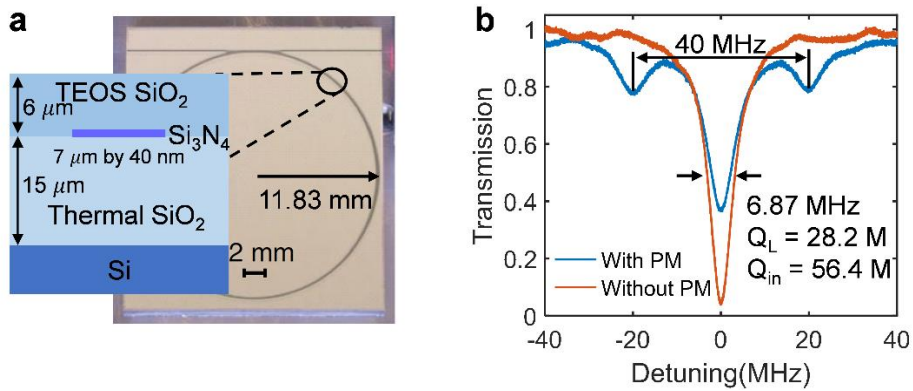

**Supplementary Fig. 1 | Integrated resonator design and Q. a**, The resonator waveguide is a high-aspect-ratio design, 7 μm wide by 40 nm thick. The resonator ring radius of 11.83 mm leads to a roundtrip length of 7.4 cm. **b**, Spectral scans by a probe laser with and without sidebands separated by 40 MHz created by the phase modulation using an electro-optic modulator measures a resonator linewidth of 6.87 MHz and a 28.2 Million loaded and 56.4 Million intrinsic Q. M, Million. PM, phase modulation. TEOS-PECVD, tetraethoxysilane pre-cursor plasma-enhanced chemical vapour deposition.

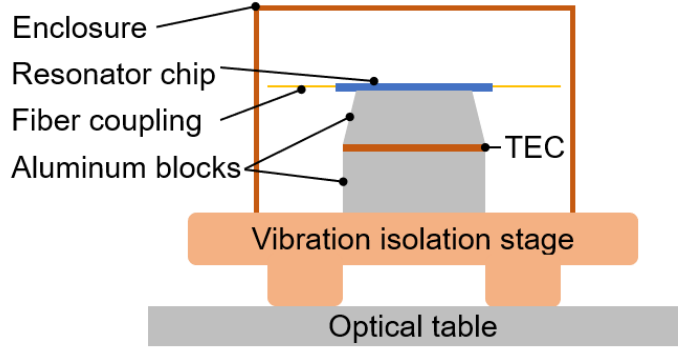

**Supplementary Fig. 2 | Resonator chip enclosure.** The resonator chip is simply enclosed by a box on a vibration isolation stage and temperature-stabilized at a mK stability by a thermo-electric cooler (TEC) connected to Vescent™ SLICE-QTC temperature controller. Two bare fibers are used to couple light in and out of the chip.

**II. PDH lock loop noise modeling.** The frequency discrimination slope using a critically coupled resonator with a certain linewidth $\delta\nu$ and a Q factor is expressed as[2],

$$\epsilon = \frac{8\sqrt{P_c P_s}}{\delta\nu}, \quad (1)$$

where $P_c$ is the optical carrier power and $P_s$ is the optical sideband power. Noise sources such as the photodetector noise and shot noise usually pose limits on the PDH lock performance. The shot noise equivalent frequency noise can be expressed as,

$$S_{sh} = \frac{\delta\nu}{4}\sqrt{\frac{h\nu}{P_c}}, \quad (2)$$

and photodetector noise equivalent frequency noise can be expressed as,

$$S_{sh} = S_{NEP}\frac{\delta\nu}{8P_c P_s}, \quad (3)$$

where $S_{NEP}$ is the noise equivalent power (NEP) of the photodetector used in the PDH lock. In our PDH lock experiments, the sideband power is ~0.05 of the carrier power with a total optical power of ~0.1 mW and the Thorlabs PDB450C photodetector with a conversion gain of 1 kV W$^{-1}$ and an NEP of 70 pW ($\sqrt{Hz}$)$^{-1}$ is used.

**III. OFD and SLS frequency noise measurements.** The optical frequency discriminator (OFD) is made of a fibre-based unbalanced Mach-Zehnder interferometer (MZI) with a free spectral range (FSR) of 1.03 MHz and the balanced photodetector (Thorlabs PDB450C). The frequency noise measurement using this OFD is referred to as the OFD frequency measurement. The power spectral density of the detector output, $S_{PDB}(\nu)$ in (V$^2$ Hz$^{-1}$), relates to frequency noise of the laser, $S_f(\nu)$ in (Hz$^2$ Hz$^{-1}$), by[1]:

$$S_f(\nu) = S_{PDB}(\nu)\left(\frac{\nu}{\sin(\pi\nu\tau_D)V_{PP}}\right)^2 \quad (4)$$

In the OFD frequency noise measurement, a ramp signal is applied to the fiber stretcher in the MZI to measure the peak-to-peak voltage, $V_{pp}$, and then without the ramp signal a high speed oscilloscope (InfiniiVision DSOX6004A) samples the output voltage with three different sampling speeds (10 kSa s$^{-1}$, 1 Msa s$^{-1}$, 1 Gsa s$^{-1}$). $1/\tau_D$ is the MZI FSR. 16 traces are taken for each sampling rate and a single-sided power spectral density is calculated using the 16 time traces, which is converted into the frequency noise spectrum using Eq. (4). Three frequency noise traces with different sampling rates are stitched together as a single frequency noise measurement.

The commercially available Stable Laser System™ uses an ultra-low expansion cavity with a finesse of ~400000 and a linewidth of several kHz to stabilize the single-frequency Rock™ fiber laser to provide a stability below 10$^{-15}$ at 1 s averaging time scale, which is referred to as the stable reference laser (SRL). The beatnote between the Brillouin laser and the SRL is detected by a high speed Thorlabs DET01CFC photodetector. The beatnote frequency is around 100 MHz. We employ the Keysight 53200A frequency counter to read the beatnote frequency to read two data traces with two different gate times or sampling rates (0.01 ms and 1 ms). Each trace for each gate time has 10000 sample points. The single-sided power spectral density of the time traces of the beatnote frequency is the measured frequency noise of the laser and we stitch the two frequency noise traces for a complete measurement. Such a frequency noise measurement using the SRL is referred to as the SRL frequency noise measurement.

Because of the excess frequency noise at below 1 kHz frequencies from the MZI in the OFD measurement and the unsuppressed laser noise at above 10 kHz frequencies from the SRL, as shown in Supplementary Fig. 3, we crop and stitch the OFD and SRL measurements for the complete laser frequency noise measurements where the OFD measures the far-from-carrier (below 1 kHz offset) noise and the SRL measures the close-to-carrier (above 1 kHz offset) noise. The fundamental linewidth, which is also referred to as the Schawlow-Townes linewidth, is calculated by multiplying the white-frequency-noise floor by π, which characterizes the short-term linewidth of the laser and is an intrinsic property of the laser determined by the fundamental physical fluctuations, while the integral linewidth calculated by integrating the single sided phase noise from the highest measured frequency offset (10MHz) to the frequency offset at which the integral is $\pi^{-1}$ rad$^2$ is a metric that measures the low-frequency noise and where the environmental noise plays a significant role, given by the following equations[3],

$$\Delta\nu_{ST} = \pi S_w, \tag{5}$$

$$\int_{\Delta\nu}^{\infty} \frac{S_f(\nu)}{f^2} df = \frac{1}{\pi}. \tag{6}$$

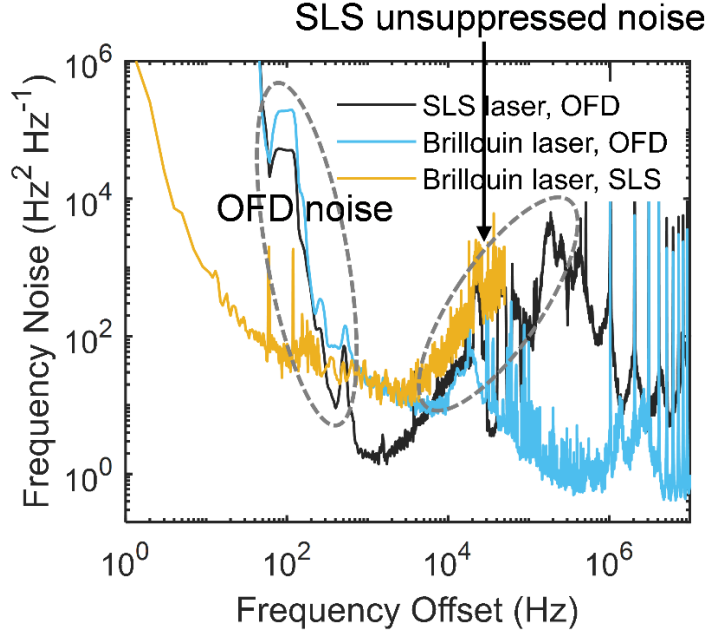

**Supplementary Fig. 3 | Complete trace of the OFD and SRL frequency noise measurements.** In the optical frequency discriminator (OFD) frequency noise measurements of the stable reference laser (SRL) (black trace) and the Brillouin laser (blue trace) shows that the MZI's noise dominates at frequencies below ~1 kHz. In the OFD measurement of the Brillouin laser, the OFD noise shows up at frequency offset below ~1 kHz. The unsuppressed laser noise of the SRL at frequency offset above ~1 kHz is revealed by the OFD measurement of the SRL (black trace), which also shows up the SRL frequency noise measurement of the Brillouin laser (yellow trace).

**IV. Absorption loss measurement and photothermal noise modeling.** Transmission around resonance below 1 indicates the power dissipation in the resonator: $P_{disp} = P_{in}(1 - T)$. Part of the dissipated power is absorbed and converted into heat: $P_{abs} = \xi P_{disp}$, where $\xi$ is absorption loss fraction and absorption loss rate can be expressed as $\gamma_{abs} = \xi \gamma_{in}$. Since only the waveguide is heated and the 1 mm thick Si substrate remains mostly undisturbed, the thermal refractive effect dominates and thermal expansion effect is negligible. Using the thermo-optic coefficients of $SiO_2$ ($0.95 \times 10^{-5}$ K$^{-1}$) and SiN ($2.45 \times 10^{-5}$ K$^{-1}$) at 1550 nm reported in the literature[4,5], we perform a Comsol simulation that simulates the thermal heating due to absorption heating and estimates the redshift given an absorption power: $\delta f_{res} = \alpha P_{abs}$. The simulation suggests $R_{th}$= 4.98 K W$^{-1}$, $\delta f_{res}/\delta T$= 1.23 GHz K$^{-1}$, and $\alpha = \delta f_{res}/P_{abs}$= 6.11 MHz (mW)$^{-1}$. With the photothermally induced resonance shift, the resonator transmission $T$ around resonance incorporate the resonator shift that is a function of the transmission $T$ given by the following equation,

$$T = 1 - \frac{\gamma_{in}\gamma_{ex}}{[\Delta\omega - 2\pi f_D(1 - T)]^2 + (\gamma_{in} + \gamma_{ex})^2/4}, \tag{5}$$

where $f_D = \xi \alpha P_{in}$ is the parameter to be extracted by fitting the skewed resonance. Supplementary Fig. 4a shows the resonance redshift with different on-chip power levels and Supplementary Fig. 4b plots the fitted $f_D$ versus $P_{in}$ and the fitting yields $\xi\alpha = 0.647$ MHz (mW)$^{-1}$. Therefore, the absorption loss fraction is measured to be 10.6% and the corresponding absorption loss is 0.05 dB m$^{-1}$.

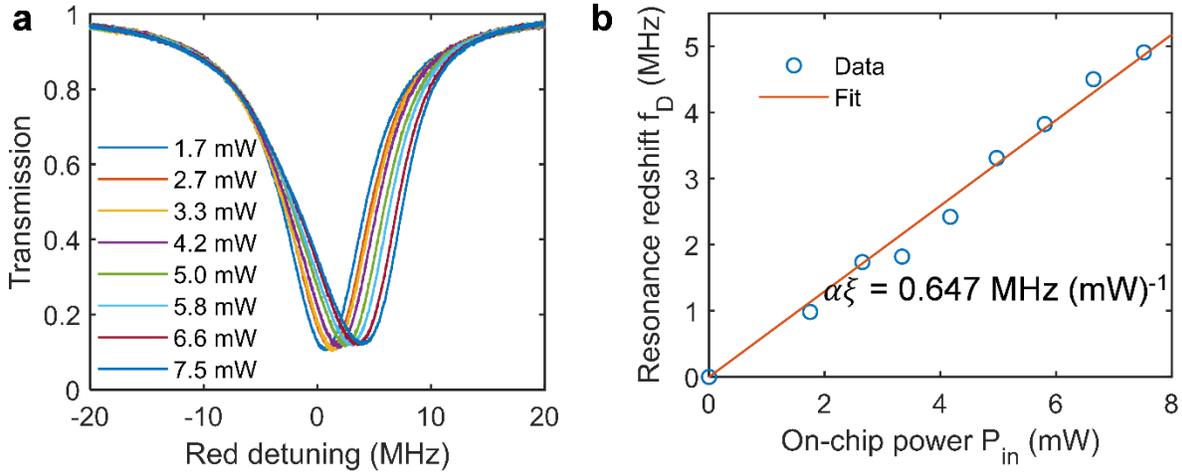

**Supplementary Fig. 4 | Resonator waveguide absorption loss measurement. a**, Spectral scan of the resonator resonance with different on-chip powers, showing the resonator redshift induced by the absorption heating effect. **b**, Resonance redshift versus the on-chip power shows a good linear fitting with a coefficient of 0.647 MHz (mW)$^{-1}$.

The relative intensity noise (RIN) of the pump laser for the Brillouin laser and the Brillouin laser first order Stokes (S1) emission is measured using a Thorlabs DET01 high speed photodetector, as shown in Supplementary Fig. 5, and is used to estimate the photothermal noise for the free-running and stabilized laser. The photothermal noise is calculated by assuming the Gaussian optical mode profile with radii of $\sigma_x$ and $\sigma_y$ and integrating all thermal mode excitation with $q$ from 0 to ∞,

$$S_f^{PT}(\omega) = S_{RIN}(\omega)(\frac{df}{dT})^2(\frac{\alpha P_{cav}}{2\pi\rho C})|\int_0^\infty \frac{q}{Dq^2+i\omega}e^{-\frac{q^2}{4}(\sigma_x^2+\sigma_y^2)}I_0[\frac{q^2}{2}(\sigma_x^2-\sigma_y^2)]dq|^2, \quad (6)$$

where the measured $df/dT$ is 1.23 GHz K$^{-1}$, the measured absorption loss $\alpha$ is 0.05 dB m$^{-1}$, $\rho$ is the density, $C$ the heat capacity, $D = \kappa/\rho C$ the thermal diffusion constant, $\kappa$ the thermal conductivity.

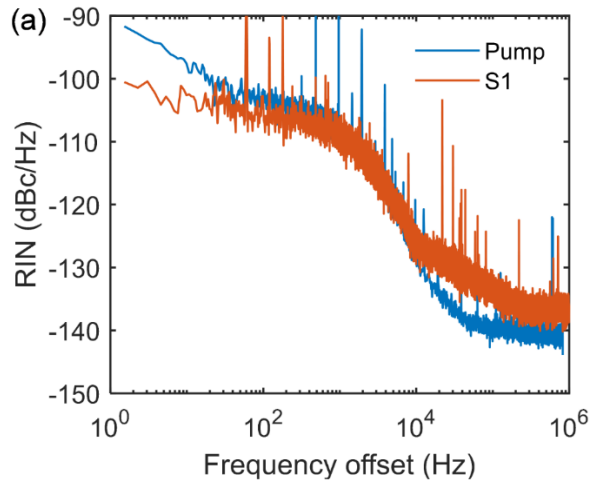

**Supplementary Fig. 5 | Laser RIN.** Relative intensity noise (RIN) of the pump laser and Brillouin laser first order Stokes (S1) is measured and used for the photothermal noise modeling.